\documentclass[lettersize,journal]{IEEEtran}
\usepackage{amsmath,amsfonts,amsthm}

\usepackage{algorithmic}
\usepackage{algorithm}
\usepackage{array}
\usepackage{subfigure}
\usepackage{textcomp}
\usepackage{stfloats}
\usepackage{url}
\usepackage{verbatim}
\usepackage{graphicx}
\graphicspath{{image/}}
\usepackage{cite}
\usepackage{booktabs}
\begin{document}

\title{Attention-based CNN-LSTM and XGBoost hybrid model for stock prediction}

\author{Zhuangwei Shi, Yang Hu, Guangliang Mo, Jian Wu
\thanks{This work was supported by the Orient Securities Co. Ltd, and the Entropy-controlled Research Project of Shenzhen Wukong Investment Management Co. Ltd. Correspondence author: Zhuangwei Shi (email:zwshi@mail.nankai.edu.cn)}
\thanks{Zhuangwei Shi and Yang Hu are with the College of Artificial Intelligence, Nankai University, Tianjin, China, and the Shenzhen Research Institute, Nankai University, Shenzhen, Guangdong, China.}
\thanks{Guangliang Mo is with the Orient Securities Co. Ltd, Shanghai, China.}
\thanks{Jian Wu is with the Research Center of Pearl River-West River Economic Belt (PWEB) Development and Finance, and the Institute of Regional Development, Guangxi Academy of Social Science, Nanning, Guangxi, China.}}

\markboth{Journal of \LaTeX\ Class Files,~Vol.~14, No.~8, August~2021}%
{Shell \MakeLowercase{\textit{et al.}}: A Sample Article Using IEEEtran.cls for IEEE Journals}


\maketitle

\begin{abstract}
    Stock market plays an important role in the economic development. Due to the complex volatility of the stock market, the research and prediction on the change of the stock price, can avoid the risk for the investors. The traditional time series model ARIMA can not describe the nonlinearity, and can not achieve satisfactory results in the stock prediction. As neural networks are with strong nonlinear generalization ability, this paper proposes an attention-based CNN-LSTM and XGBoost hybrid model to predict the stock price. The model constructed in this paper integrates the time series model, the Convolutional Neural Networks with Attention mechanism, the Long Short-Term Memory network, and XGBoost regressor in a non-linear relationship, and improves the prediction accuracy. The model can fully mine the historical information of the stock market in multiple periods. The stock data is first preprocessed through ARIMA. Then, the deep learning architecture formed in pretraining-finetuning framework is adopted. The pre-training model is the Attention-based CNN-LSTM model based on sequence-to-sequence framework. The model first uses convolution to extract the deep features of the original stock data, and then uses the Long Short-Term Memory networks to mine the long-term time series features. Finally, the XGBoost model is adopted for fine-tuning. The results show that the hybrid model is more effective and the prediction accuracy is relatively high, which can help investors or institutions to make decisions and achieve the purpose of expanding return and avoiding risk. Source code is available at \url{https://github.com/zshicode/Attention-CLX-stock-prediction}.
\end{abstract}


\section{Introduction}
\label{sec:introduction}

\IEEEPARstart{S}{tock} market plays an important role in the economic development. Due to the high return characteristics of stocks, the stock market has attracted more and more attention from institutions and investors. However, due to the complex volatility of the stock market, sometimes it will bring huge loss to institutions or investors. Considering the risk of the stock market, the research and prediction on the change of the stock price can avoid the risk for the investors.

The traditional time series model ARIMA can not describe the nonlinear time series, and needs to satisfy many preconditions before modeling, and can not achieve remarkable results in the stock forecasting. In recent years, with the rapid development of artificial intelligence theory and technology, more and more researchers apply artificial intelligence method to the financial market. On the other hand, the sequence modeling problem, focusing on natural language sequences, protein sequences, stock price sequences, and so on, is important in the field of artificial intelligence research \cite{shi2021vgaelda,jin2021lpigac}. The most representative artificial intelligence method is neural networks, which are with strong nonlinear generalization ability.

Recurrent Neural Network (RNN) was adopted for analyzing sequential data via neural network architecture, and Long Short-Term Memory (LSTM) model is the most commonly used RNN. LSTM introduced gate mechanism in RNN, which can be seen as simulation for human memory, that human can remember useful information and forget useless information \cite{jin2021chinese}. Attention Mechanism \cite{transformer2017,jin2022nimgsa} can be seen as simulation for human attention, that human can pay attention to useful information and ignore useless information. Attention-based Convolutional Neural Networks (ACNN) are widely used for sequence modeling \cite{JIN2021265,lin2022dattprot}. Combining Attention-based Convolutional Neural Networks and Long Short-Term Memory, is a self-attention based sequence-to-sequence (seq2seq) \cite{seq2seq2014} model to encode and decode sequential
data. This model can solve long-term dependency problem in LSTM, hence, it can better model long sequences. LSTM can capture particular long-distance correspondence that fits the sturcture of LSTM itself, while ACNN can capture both local and global correspondence. Therefore, this architecture is more flexible and robust.

Transformer \cite{transformer2017} is the most successful sequential learning self-attention based model. Experiments on natural language processing demonstrates that Transformer can better model long sequences. Bidirectional Encoder Representation Transformer (BERT) with pretraining \cite{Devlin2018BERT} can perform better than the basic Transformer. Pretraining  is a method to significantly improve the performance of Transformer (BERT).

This paper proposes a hybrid deep learning model to predict the stock price. Different from the traditional hybrid prediction model, the proposed model integrates the time series model ARIMA and the neural networks in a non-linear relationship, which combines the advantages of the two vanilla models, and improves the prediction accuracy. The stock data is first preprocessed through ARIMA. The stock sequence is put into neural networks (NN) or XGBoost after preprocessing via ARIMA(p=2,q=0,d=1). Then, the deep learning architecture formed in pretraining-finetuning framework \cite{Devlin2018BERT,jin2022tlcrys} is adopted. The pre-training model is the Attention-based CNN-LSTM model based on sequence-to-sequence framework, where the Attention-based CNN is encoder, and the Bidirectional LSTM is decoder. The model first uses convolution to extract the deep features of the original stock data, and then uses the Long Short-Term Memory networks to mine the long-term time series features. Finally, the XGBoost model is adopted for fine-tuning, which can fully mine the information of the stock market in multiple periods. Our proposed Attention-based CNN-LSTM and XGBoost hybrid model is so called AttCLX.

The results show that the model is more effective and the prediction accuracy is relatively high, which can help investors or institutions to make decisions and achieve the purpose of expanding returns and avoiding risks. The source code of this paper is available at \url{https://github.com/zshicode/Attention-CLX-stock-prediction}. We conduct empirical study on the stock price of Back of China (601988.SH) in Chinese stock market. The data is downloaded from Tushare(\url{www.tushare.pro}). The stock price data on Tushare is with public availability.

\section{Materials and Methods}

\subsection{ARIMA}

Classical stock prediction methods are based on ARMA (Auto Regressive Moving Average) model and ARIMA (Auto Regressive Integrated Moving Average) model. An ARMA(p,q) model
\begin{equation}
    s_t=a_0+\sum_{i=1}^p a_is_{t-i}+w_t+\sum_{i=1}^q b_iw_{t-i},
\end{equation}
where a,b are parameters, w is noise. ARMA model can be used when sequence $s_{1:N}$ is stationary, which means
\begin{equation}
    \mathbb{E}[s_t]=\mathrm{Constant}.
\end{equation}
\begin{equation}
    \mathrm{Cov}(s_t,s_{t-k})=\mathrm{Constant}.
\end{equation}
where $t=1,2,...,N$ and $k=1,2,...,t$.

When sequence is non-stationary, ARIMA(p,q,d) adopts $d$-order difference to the sequence. In stock prediction, the first-order difference $x_{1:N}=\dot s_{1:N}$ (i.e. $x_k=s_k-s_{k-1}$) is usually considered as stationary sequence.

We conduct empirical study on the stock price of Back of China (601988.SH) in Chinese stock market. The data is downloaded from Tushare(\url{www.tushare.pro}). The stock price data on Tushare is with public availability. The data is selected from the data from January 1, 2007 to March 31, 2022, the data in one day denotes a point of the sequence.

\begin{figure}
    \centering
    \includegraphics[width=0.46\textwidth]{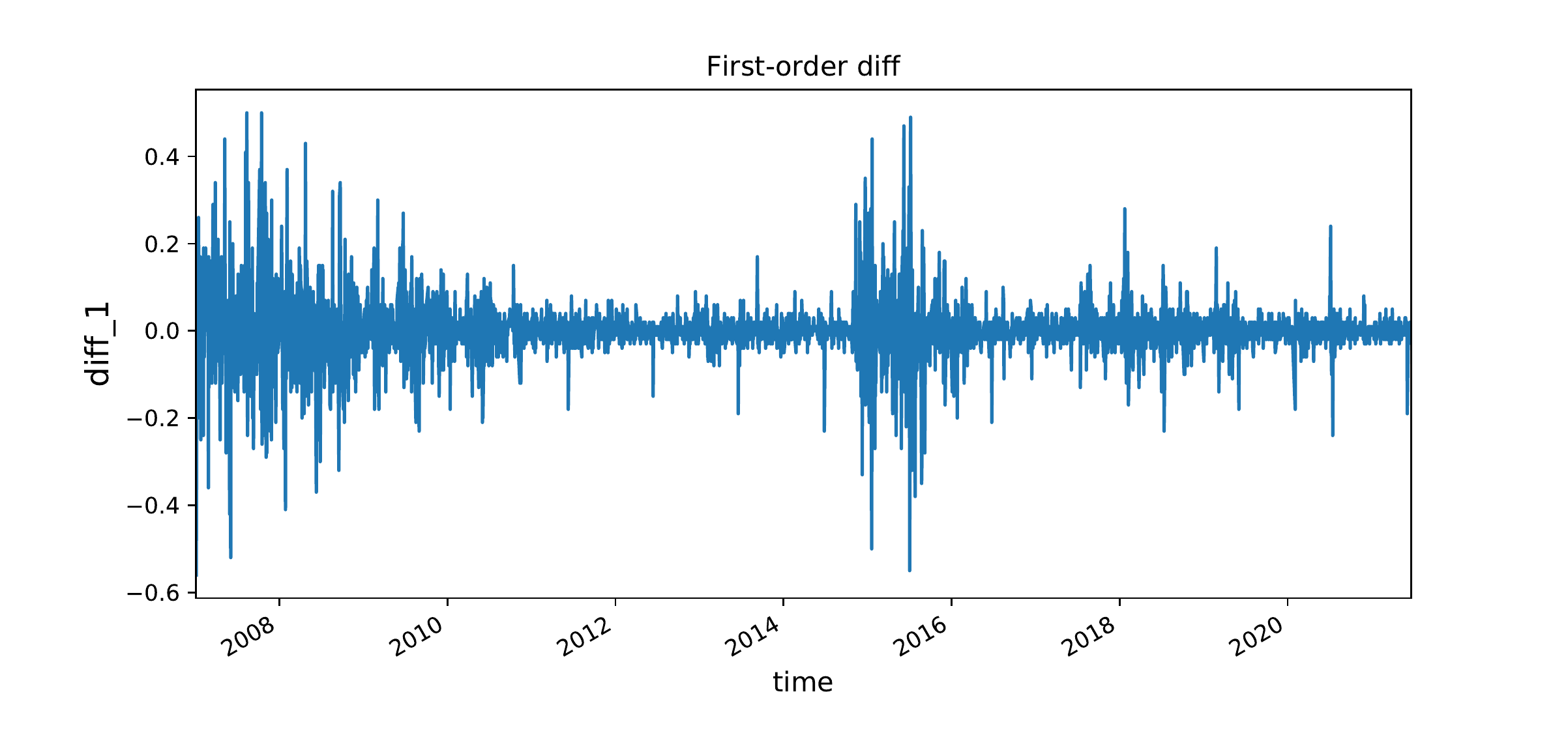}
    \caption{The first-order difference.}\label{fig:d}
\end{figure}

\begin{figure}
    \centering
    \includegraphics[width=0.46\textwidth]{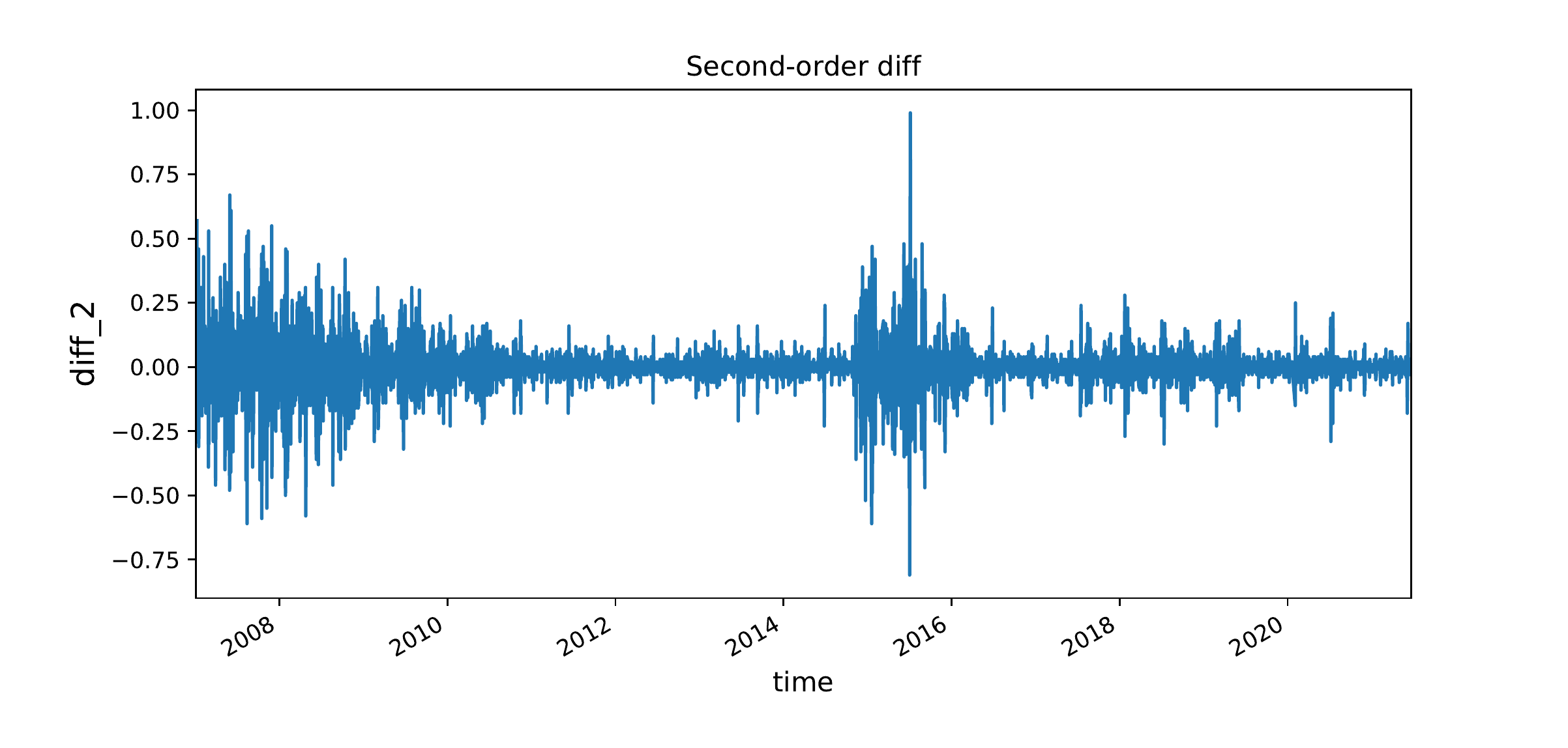}
    \caption{The second-order difference.}\label{fig:dd}
\end{figure}

ADF test is adopted for testing the stationary condition of time series. Adopting ADF test for the original sequence and first-order difference sequence. The results are in Table \ref{tab:adf} and \ref{tab:adf1}. When p-value is more than 0.562 or Critical Value (1\%) is more than -3.44, the sequence is non-stationary. The ADF test shows that the original sequence is non-stationary and first-order difference sequence is stationary. The first-order difference sequence and second-order difference sequence are shown on Fig.\ref{fig:d} and \ref{fig:dd}.

\begin{table}[]
    \caption{ADF test for original sequence}
    \label{tab:adf}
    \centering
    \begin{tabular}{@{}cc@{}}
    \toprule
    Metric & Value      \\ \midrule
    Test Statistic Value    &     -2.35539\\
    p-value                 &     0.154726\\
    Lags Used               &           16\\
    Number of Observations Used  &    3484\\
    Critical Value(1\%)     &      -3.43223\\
    Critical Value(5\%)     &      -2.86237\\
    Critical Value(10\%)    &      -2.56721\\ \bottomrule
    \end{tabular}
\end{table}

\begin{table}[]
    \caption{ADF test for first-order diff sequence}
    \label{tab:adf1}
    \centering
    \begin{tabular}{@{}cc@{}}
    \toprule
    Metric & Value      \\ \midrule
    Test Statistic Value &           -14.7498\\
    p-value              &        2.49565e-27\\
    Lags Used            &                 15\\
    Number of Observations Used   &      3484\\
    Critical Value(1\%)  &            -3.43223\\
    Critical Value(5\%)  &            -2.86237\\
    Critical Value(10\%) &            -2.56721\\ \bottomrule
    \end{tabular}
\end{table}

\subsection{Deep learning on sequential data}

\begin{figure}
    \centering
    \includegraphics[width=0.3\textwidth]{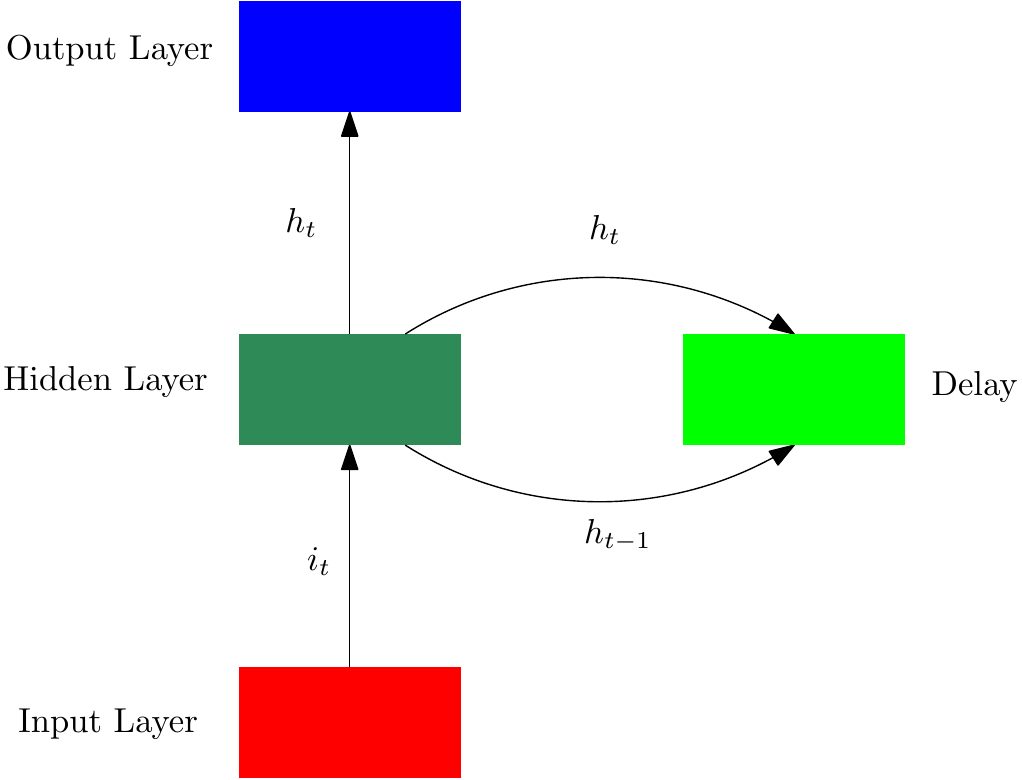}
    \caption{RNN unit.}
    \label{fig:rnn}
\end{figure}

In basic feed-forward neural network (FFNN), output of current moment $o_t$ is only determined by input of current moment $i_t$, which suppress the ability of FFNN to  model time-series data. In recurrent neural network (RNN), a delay is used to save the latent state of latest moment $h_{t-1}$, then, latent state of current moment $h_t$ is determined by both $h_{t-1}$ and $i_t$.  See Fig. \ref{fig:rnn}. \cite{lstm1997} suggested that RNN may vanish the gradient as error propagates through time dimension, which leads to long-term dependency problem. Human can selectively remember information. Through gated activation function, LSTM (long short-term memory) model can selectively remember updated information and forget accumulated information.

Sequence-to-sequence (seq2seq) \cite{seq2seq2014} model adopted autoencoder (i.e. encoder-decoder architecture) for analyzing sequential data. Sequence-to-sequence model (seq2seq) \cite{seq2seq2014} is constructed through an encoder-decoder architecture, which enhances the ability of LSTM to learn hidden information through data with noise. In seq2seq, the encoder is an LSTM that encodes the inputs into the context (commonly the hidden state at last $h_N$), then decode the context in the decoder. In the decoder, output at previous moment is input at next moment.

To optimize the decoded sequence, beam search \cite{beamsearch2016} was adopted in seq2seq model. Both beam search and Viterbi algorithm in hidden Markov model (HMM) are based on dynamic programming. Solving optimal estimation of current state according to observation and previous state, is called decoding or inference in HMM. Solving $p(x_k|y_{1:k})$ and $p(x_k|y_{1:N})$ respectively, is equivalent to forward-backward algorithm in HMM. The optimal bidirectional estimation can be obtained through the distribution of $x_k$. That is the probabilistic perspective that the bidirectional LSTM which combines forward and backward, is proposed \cite{jin2021chinese}.

\subsection{Attention mechanism}

Human usually pays attention to salient information. Attention mechanism is a technique in deep learning based on human cognitive system. For input $X=(x_1,x_2,...,x_N)$, give query vector $q$, depict the index of selected information by attention $z=1,2,...,N$, then the distribution of attention
\begin{equation} 
\alpha_{i} =p(z=i | X, \mathbf{q}) =\frac{\exp \left(s\left(\mathbf{x}_{i}, \mathbf{q}\right)\right)}{\sum_{j=1}^{N} \exp \left(s\left(\mathbf{x}_{j}, \mathbf{q}\right)\right)}.
\end{equation}
i.e.
\begin{equation}
\alpha_{i}=\mathrm{softmax}\left(s\left(\mathbf{x}_{i}, \mathbf{q}\right)\right) 
\end{equation}
Here
\begin{equation} 
s\left(\mathbf{x}_{i}, \mathbf{q}\right)=\frac{\mathbf{x}_{i}^{\mathrm{T}} \mathbf{q}}{\sqrt{d}}
\end{equation}
is attention score through scaled dot product, $d$ is the dimension of input. Suppose the input key-value pairs $(K, V)=\left[\left(\mathbf{k}_{1}, \mathbf{v}_{1}\right), \cdots,\left(\mathbf{k}_{N}, \mathbf{v}_{N}\right)\right]$, for given $q$, attention function
\begin{equation} 
\mathrm{att}((K, V), \mathbf{q}) =\sum_{i=1}^{N} \alpha_{i} \mathbf{v}_{i} =\sum_{i=1}^{N} \frac{\exp \left(s\left(\mathbf{k}_{i}, \mathbf{q}\right)\right)}{\sum_{j} \exp \left(s\left(\mathbf{k}_{j}, \mathbf{q}\right)\right)} \mathbf{v}_{i} .
\end{equation}

Multi-head mechanism is usually adopted through multi-query $Q=\left[\mathbf{q}_{1}, \cdots, \mathbf{q}_{M}\right]$ for attention function computation.
\begin{equation} 
\mathrm{att}((K, V), Q)=\left(\mathrm{att}\left((K, V), \mathbf{q}_{1}\right)||\cdots||\mathrm{att}\left((K, V), \mathbf{q}_{M}\right)\right).
\end{equation}
Here, $||$ denotes Concatenate operation. This is so called multi-head attention (MHA).

Attention mechanism can be adopted to generate data-driven different weights. Here, $Q,K,V$ are all obtained through linear transform of $X$, and $W_Q,W_K,W_V$ can be adjusted dynamicly.
\begin{equation} 
Q =W_{Q} X,K =W_{K} X,V =W_{V} X.\label{eq:selfatt}
\end{equation}
This is so called self-attention. Similarly, output
\begin{equation} 
\mathbf{h}_{i} =\operatorname{att}\left((K, V), \mathbf{q}_{i}\right) .
\end{equation}
Hence
\begin{equation}
h_i=\sum_{j=1}^{N} \alpha_{i j} \mathbf{v}_{j} =\sum_{j=1}^{N} \operatorname{softmax}\left(s\left(\mathbf{k}_{j}, \mathbf{q}_{i}\right)\right) \mathbf{v}_{j} 
\end{equation}
Adopting scaled dot product score, the output
\begin{equation} 
H=V \operatorname{softmax}\left(\frac{K^{\mathrm{T}} Q}{\sqrt{d}}\right).\label{eq:selfatt:output}
\end{equation}

\subsection{Method}

\subsubsection{Preprocessing}

This paper proposes a hybrid deep learning model to predict the stock price. Different from the traditional hybrid prediction model, the proposed model integrates the time series model ARIMA and the neural networks in a non-linear relationship, which combines the advantages of the two vanilla models, and improves the prediction accuracy. 

The stock data is first preprocessed through ARIMA. The stock sequence is put into neural networks (NN) or XGBoost after preprocessing via ARIMA(p=2,q=0,d=1). Putting original stock market data into ARIMA, can output a new series that depict state more effectively.

\begin{figure}
    \centering
    \includegraphics[width=0.46\textwidth]{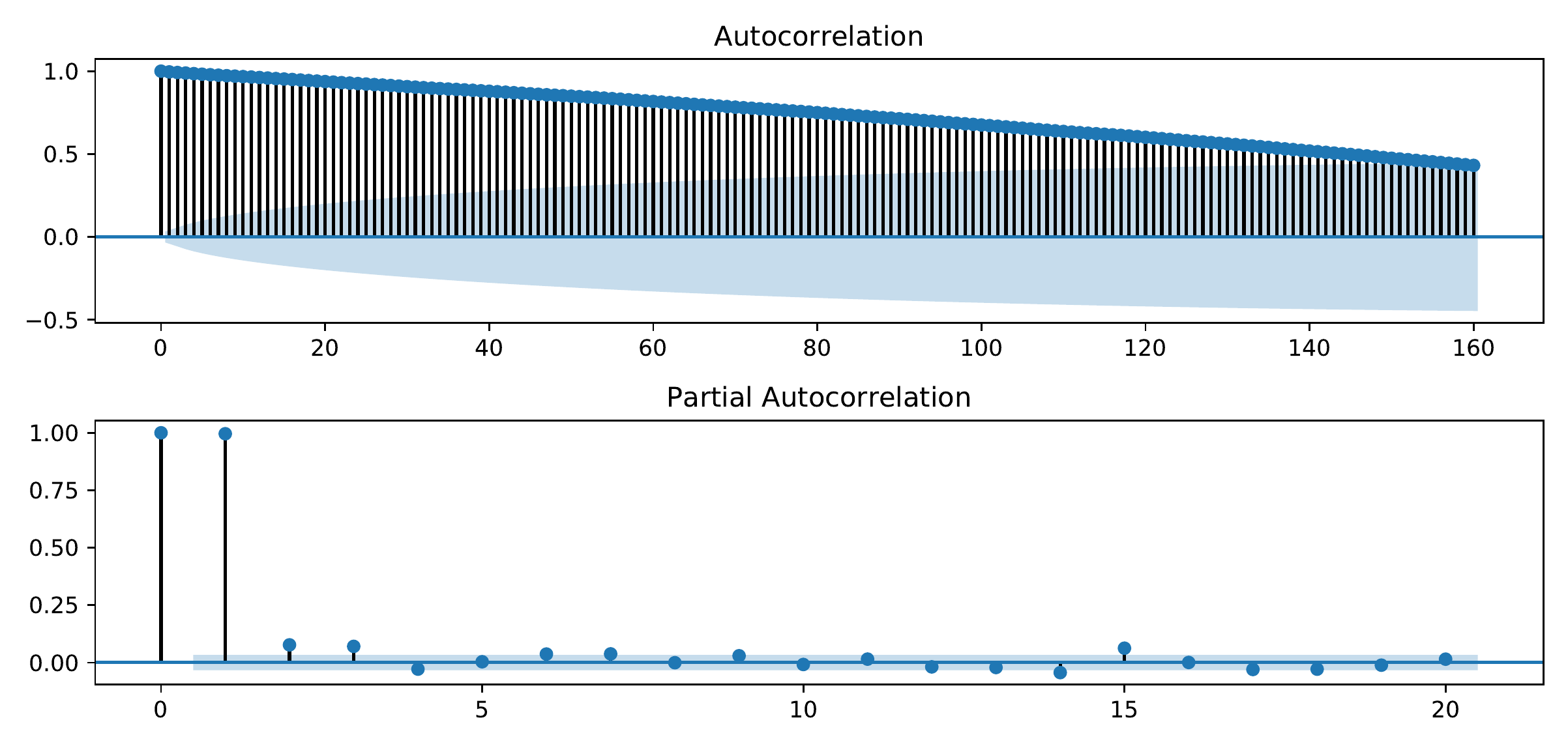}
    \caption{The ACF and PACF of the original sequence.}\label{fig:f1}
\end{figure}

\begin{figure}
    \centering
    \includegraphics[width=0.46\textwidth]{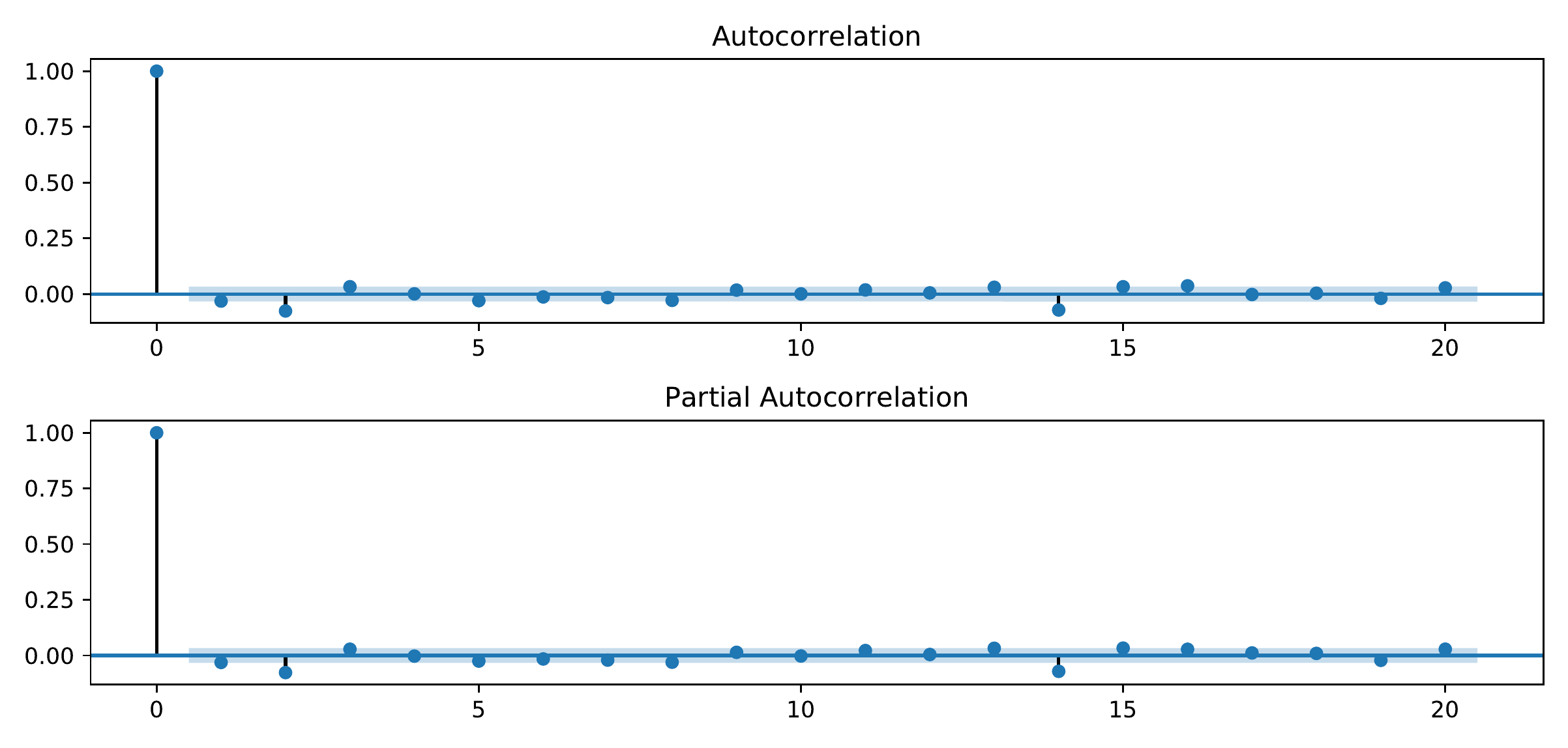}
    \caption{The ACF and PACF of the first-order difference.}\label{fig:f2}
\end{figure}

The ADF results in Table \ref{tab:adf} and \ref{tab:adf1} shows that the original sequence is non-stationary and first-order difference sequence is stationary. After determining d=1, we need to determine AR(p) and MA(q) in ARIMA. We use the Autocorrelation Figure (ACF) and Partial Autocorrelation Figure (PACF). The ACF and PACF of the original sequence and first-order difference sequence are shown on Fig. \ref{fig:f1} and \ref{fig:f2}. Fig. \ref{fig:f1} shows that PACF truncates when order=2, meaning that we should adopt AR(2). The ACF is with long tail for any order, meaning that we should adopt MA(0). Hence p=2, q=0.

\subsubsection{Pretraining}

Then, the deep learning architecture formed in pretraining-finetuning framework is adopted. The pre-training model is the Attention-based CNN-LSTM model based on sequence-to-sequence framework, where the Attention-based CNN is encoder, and the Bidirectional LSTM is decoder. The model first uses convolution to extract the deep features of the original stock data, and then uses the Long Short-Term Memory networks to mine the long-term time series features. Finally, the XGBoost model is adopted for fine-tuning, which can fully mine the information of the stock market in multiple periods. 

Seq2seq suppress the effect of noise through encoder-decoder architecture. Based on deep learning, hidden information of state is depicted more effectively, while the model would not satisfy the assumptions of linear property of stock price. LSTM receives the context from Attention-based CNN (ACNN)encoder.

The ACNN encoder block consists of self-attention layer and CNN. $Q,K,V$ are computed through Eq. \eqref{eq:selfatt} after self-attention layer, and $H$ is computed through Eq. \eqref{eq:selfatt:output}. This is the input of LSTM decoder block. Encoder-decoder layers depicts relationship between current sequence and previous sequence, and relationship between current sequence and embedding. Encoder is still with multi-head mechanism. When the $k$-th embedding is being decoded, only $k-1$-th and previous decoding can be seen. This multi-head mechanism is masked multi-head attention. 

Attention-based CNN (ACNN) can capture both global and local dependency that LSTM may not \cite{JIN2021265}, which enhance the robustness. In our proposed encoder-decoder framework, we can adopt a ACNN-LSTM structure. Attention is usually before memory in human cognitive system. The reason why ACNN can capture long-term dependency, is that it integrates multi-head self attention and convolution. Combining LSTM and ACNN can enhance both structural advantages and ability for time-series modeling. Integrating multi-head attention and multi-scale convolutional kernel, ACNN enocder can capture saliency that LSTM may not, while LSTM can better depict time-series property.

\subsubsection{Fine-tuning}

\begin{figure*}
    \centering
    \includegraphics[width=0.8\textwidth]{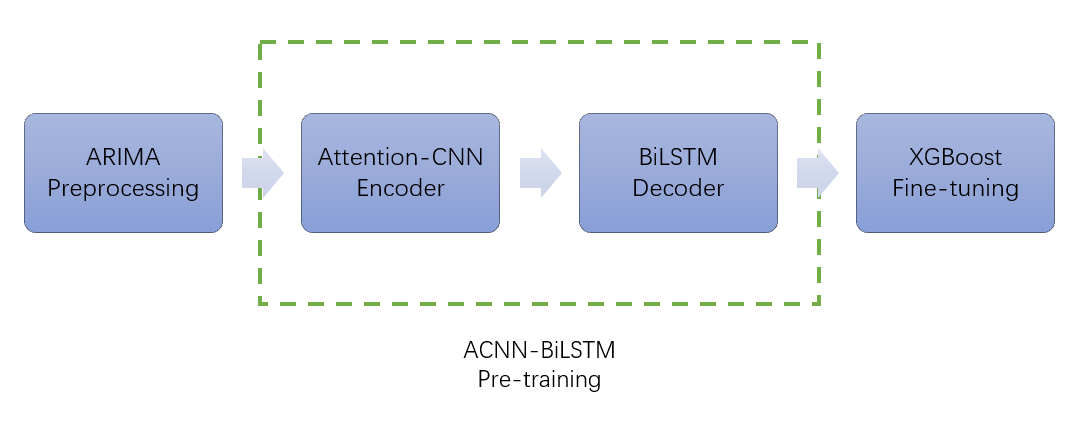}
    \caption{Attention-based CNN-LSTM and XGBoost hybrid model (AttCLX).}\label{fig:structure}
\end{figure*}

After decoding, the output is obtained through a XGBoost regressor for precise extraction of features and fine-tuning. Our proposed Attention-based CNN-LSTM and XGBoost hybrid model is so called AttCLX, which is shown on Fig. \ref{fig:structure}.

As the fine-tuning model, XGBoost \cite{2016XGBoost} is with strong expansion and flexibility. It integrates multiple tree models to build a stronger learner model. Based on the pre-training, we propose fine-tuning based on XGBoost, and establish a regression prediction model for stock data. XGBoost fine-tuning model also achieves better predictive ability and generalization ability.

\section{Experiments}

\subsection{Modification of model}

After preprocessing by ARIMA, the input of the neural networks, is a two-dimensional matrix of data at intervals of a period of time, with a size of TimeWindow$\times$Features. In Empirical studies on stock prediction, features include the basic stock market data (opening price, closing price, highest price, lowest price, trading volume, trading amount). The ARIMA-processing sequence along with the residual sequence are also concatenated as features.

We adopted look back trick for time-series forecasting, and the look back number is 20, i.e. $o_t$ can be obtained through $o_{t-1},\cdots,o_{t-5}$. This means that the TimeWindow width is 20. The layer number of LSTM is 5, and the size is 64. The epoch number is 50. Model is trained by introducing dropout \cite{2014Dropout}, and the dropout rate is 0.3. The head number is 4.

The experiments are on an NVIDIA GTX2070 GPU with 8GB memory. The model is trained through Adam optimizer \cite{2014Adam}, and learning rate is 0.01.

The data used in this article comes from the open and free public dataset in Tushare (\url{https://www.tushare.pro/}) for the research of stock market in China, which has the characteristics of rich data, simple use, and convenient implementation. It is very convenient to obtain the basic market data of stocks by calling the API.

\begin{figure}
    \centering
    \includegraphics[width=0.46\textwidth]{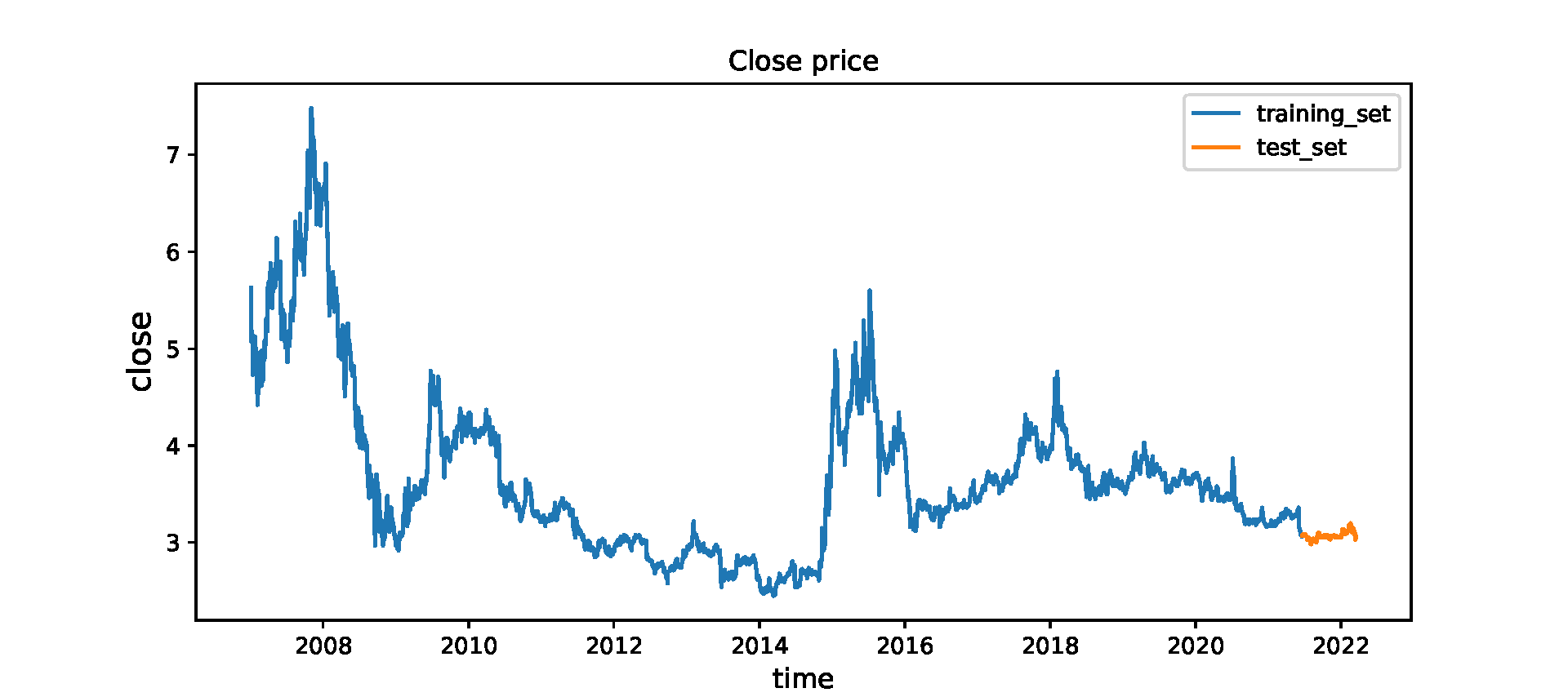}
    \caption{The train set and test set.}\label{fig:price}
\end{figure}

The implementation details of this paper can refer to source code of this paper at \url{https://github.com/zshicode/Attention-CLX-stock-prediction}. We conduct empirical study on the stock price of Back of China (601988.SH) in Chinese stock market. The data is downloaded from Tushare(\url{www.tushare.pro}). The stock price data on Tushare is with public availability. The data is selected from the data from January 1, 2007 to March 31, 2022, the data in one day denotes a point of the sequence. The train set and test set was divided on June 22, 2021, as shown on Fig. \ref{fig:price}. This means that there are 3500 training samples and 180 testing samples. The batch size is 32.

\subsection{Prediction performance}

\begin{figure}
    \centering
    \includegraphics[width=0.46\textwidth]{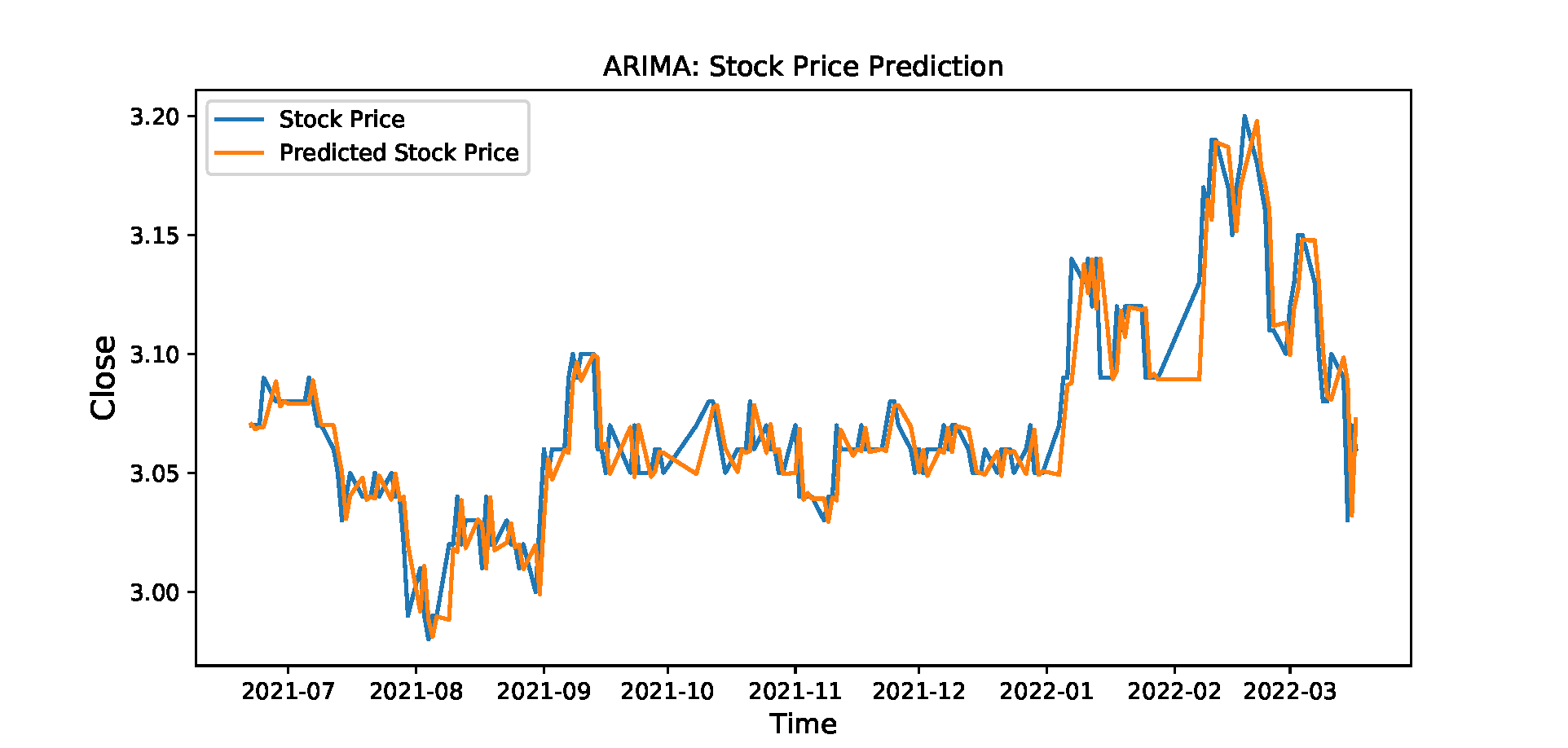}
    \caption{The ARIMA for stock price prediction.}\label{fig:arima}
\end{figure}

\begin{figure}
    \centering
    \includegraphics[width=0.46\textwidth]{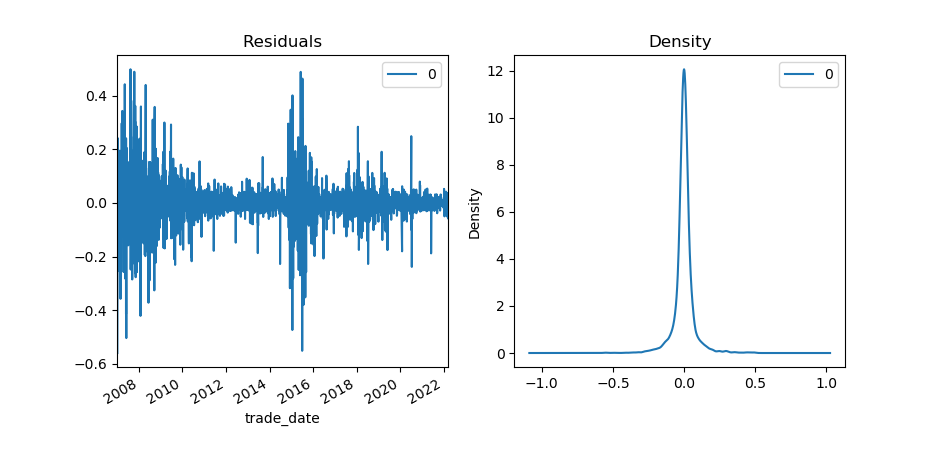}
    \caption{The resudial and residual density plot.}\label{fig:f3}
\end{figure}

The stock price prediction result of ARIMA model is shown on Fig. \ref{fig:arima}. The resudial and residual density plot are shown on Fig. \ref{fig:f3}.

\begin{figure}
    \centering
    \includegraphics[width=0.46\textwidth]{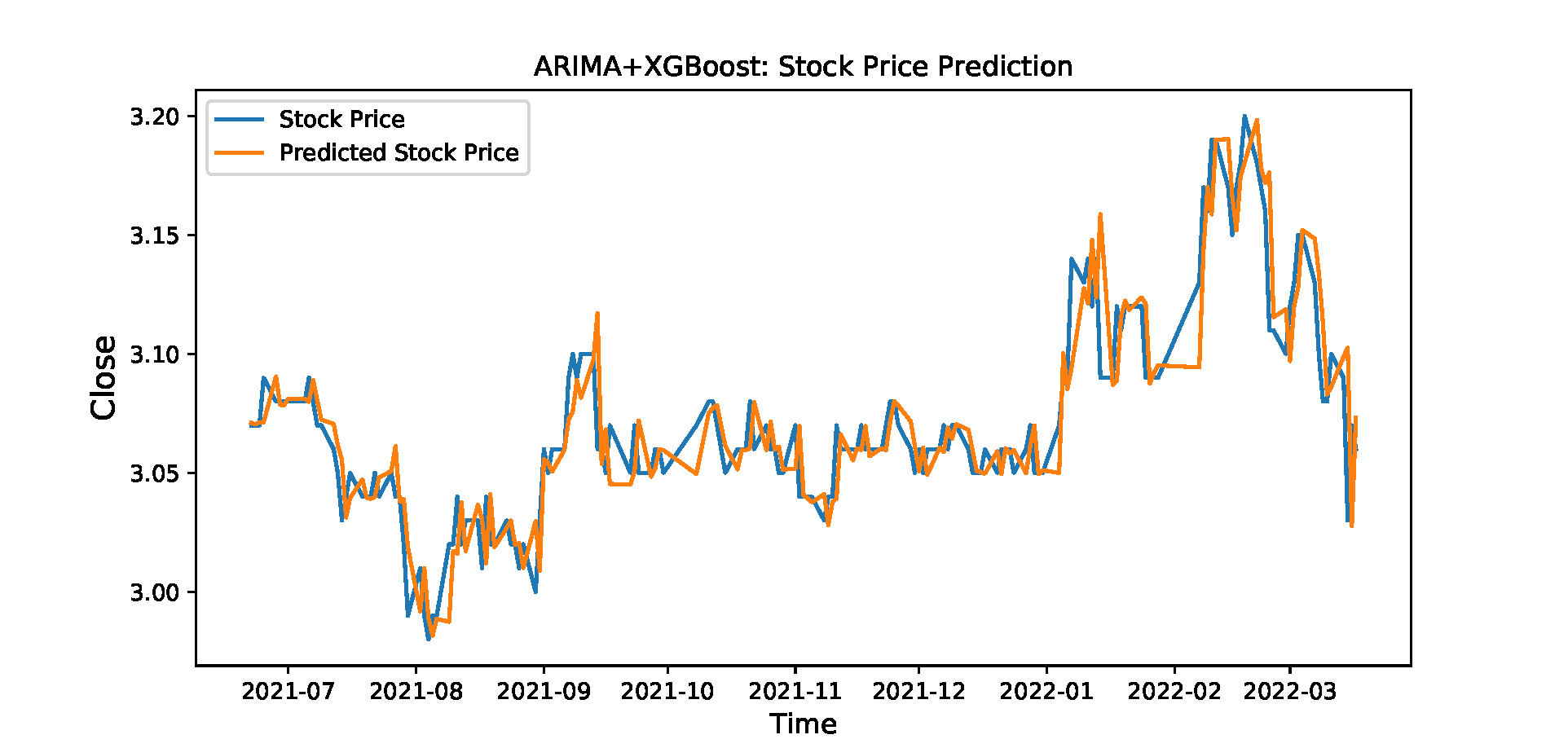}
    \caption{The ARIMA+XGBoost for stock price prediction.}\label{fig:xgbstock}
\end{figure}

\begin{figure}
    \centering
    \includegraphics[width=0.46\textwidth]{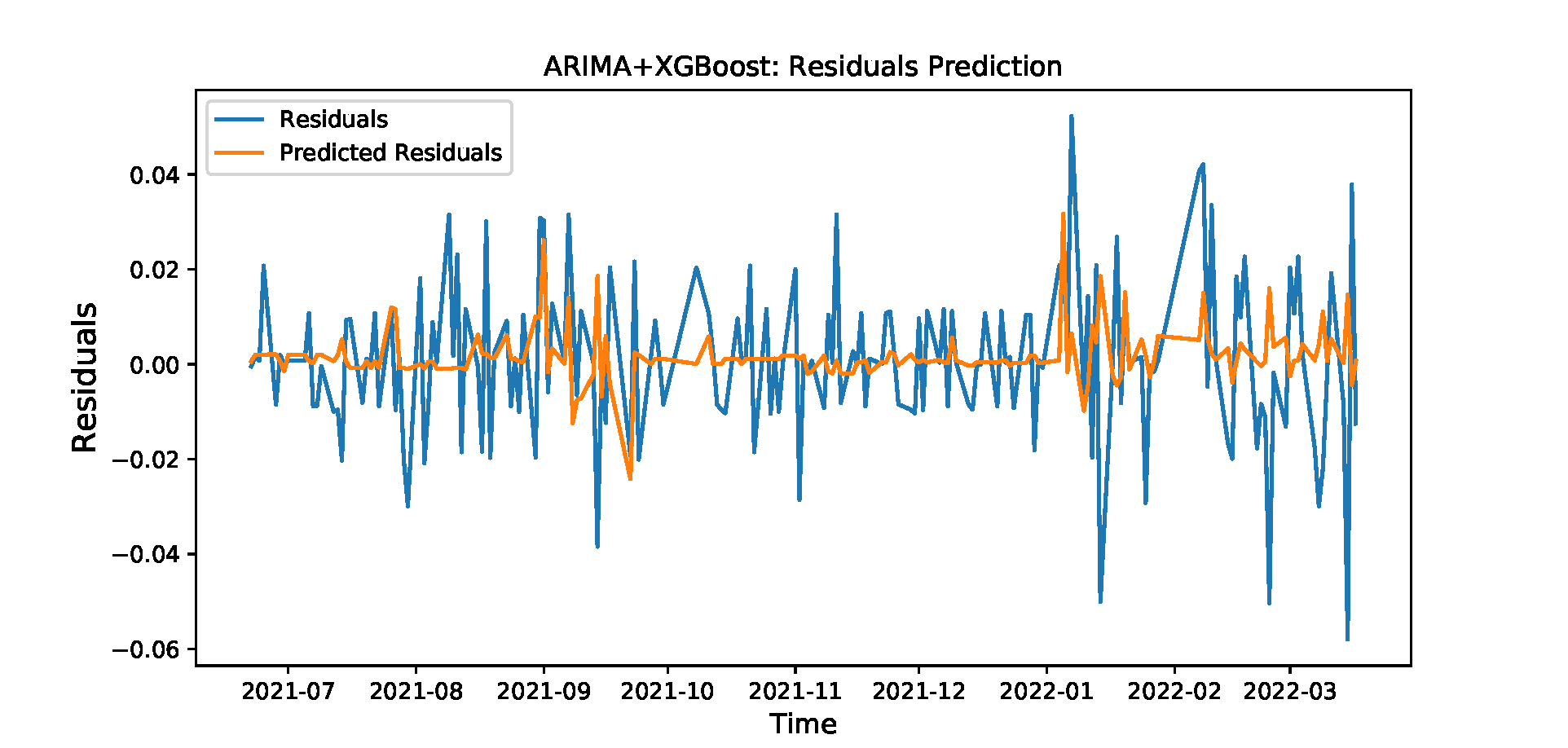}
    \caption{The ARIMA+XGBoost for resudial prediction.}\label{fig:xr}
\end{figure}

The stock price prediction result of ARIMA+XGBoost model is shown on Fig. \ref{fig:xgbstock}. The resudial and residual density plot are shown on Fig. \ref{fig:xr}.

\begin{figure}
    \centering
    \includegraphics[width=0.46\textwidth]{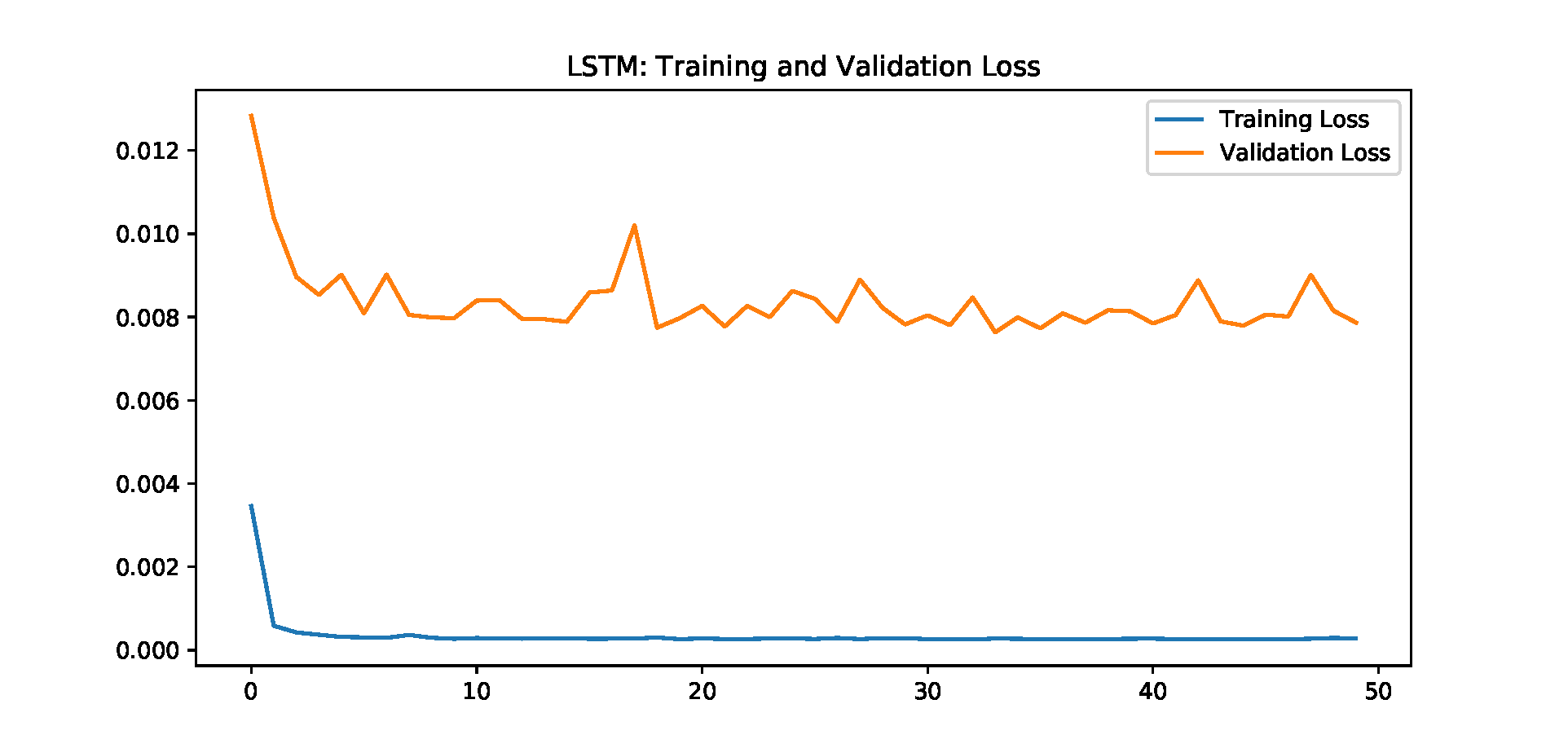}
    \caption{The loss curves of original sequence of ARIMA+SingleLSTM.}\label{fig:loss}
\end{figure}

\begin{figure}
    \centering
    \includegraphics[width=0.46\textwidth]{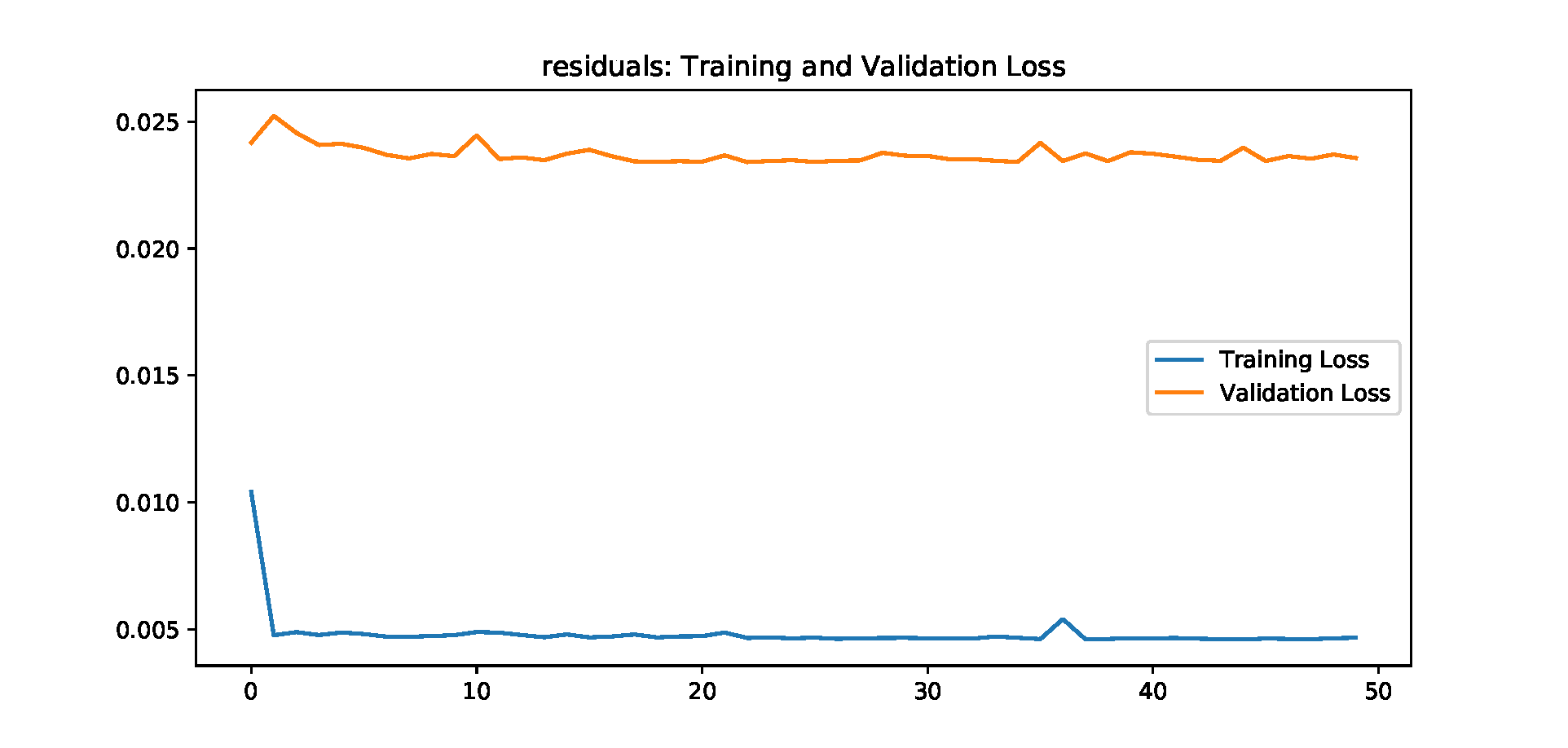}
    \caption{The loss curves of residual sequence of ARIMA+SingleLSTM.}\label{fig:lossr}
\end{figure}

The loss curves of original sequence and residual sequence of ARIMA+SingleLSTM are shown on Fig. \ref{fig:loss} and Fig. \ref{fig:lossr}.

\begin{figure}
    \centering
    \includegraphics[width=0.46\textwidth]{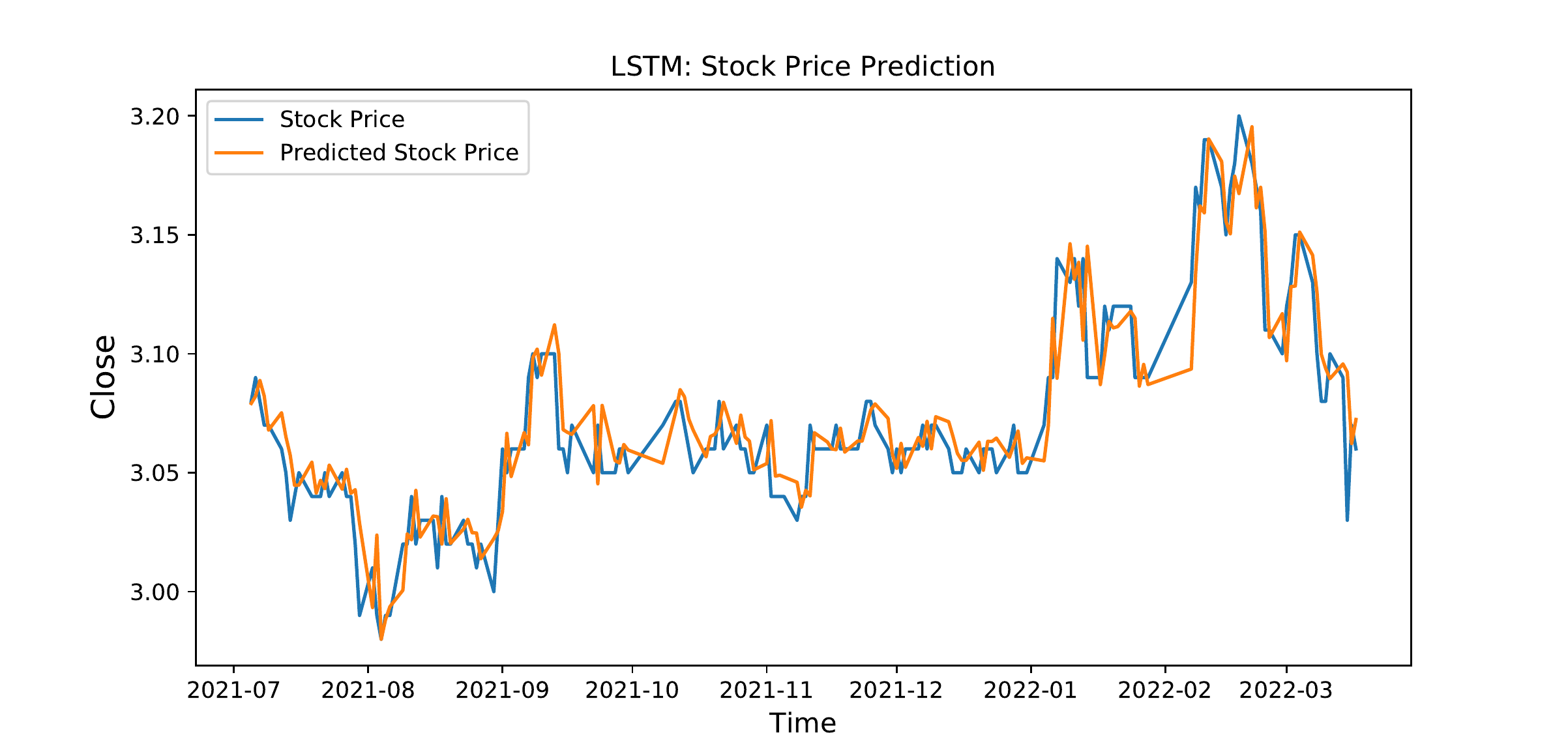}
    \caption{The stock price prediction result of ARIMA+SingleLSTM.}\label{fig:lstm}
\end{figure}

\begin{figure}
    \centering
    \includegraphics[width=0.46\textwidth]{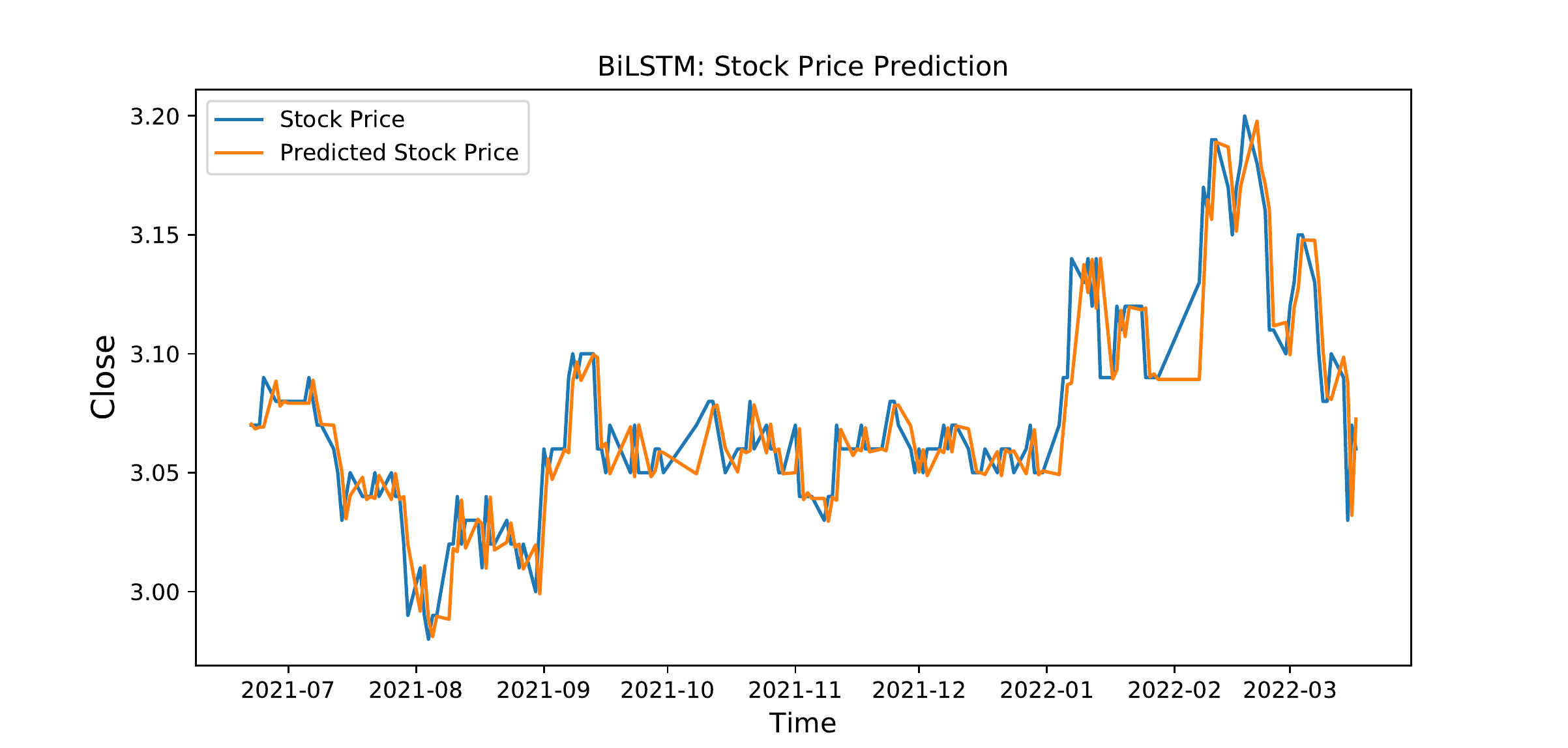}
    \caption{The stock price prediction result of ARIMA+BiLSTM.}\label{fig:bilstm}
\end{figure}

The stock price prediction result of ARIMA+SingleLSTM model and ARIMA+BiLSTM model are shown on Fig. \ref{fig:lstm} and Fig. \ref{fig:bilstm}.

\begin{figure}
    \centering
    \includegraphics[width=0.46\textwidth]{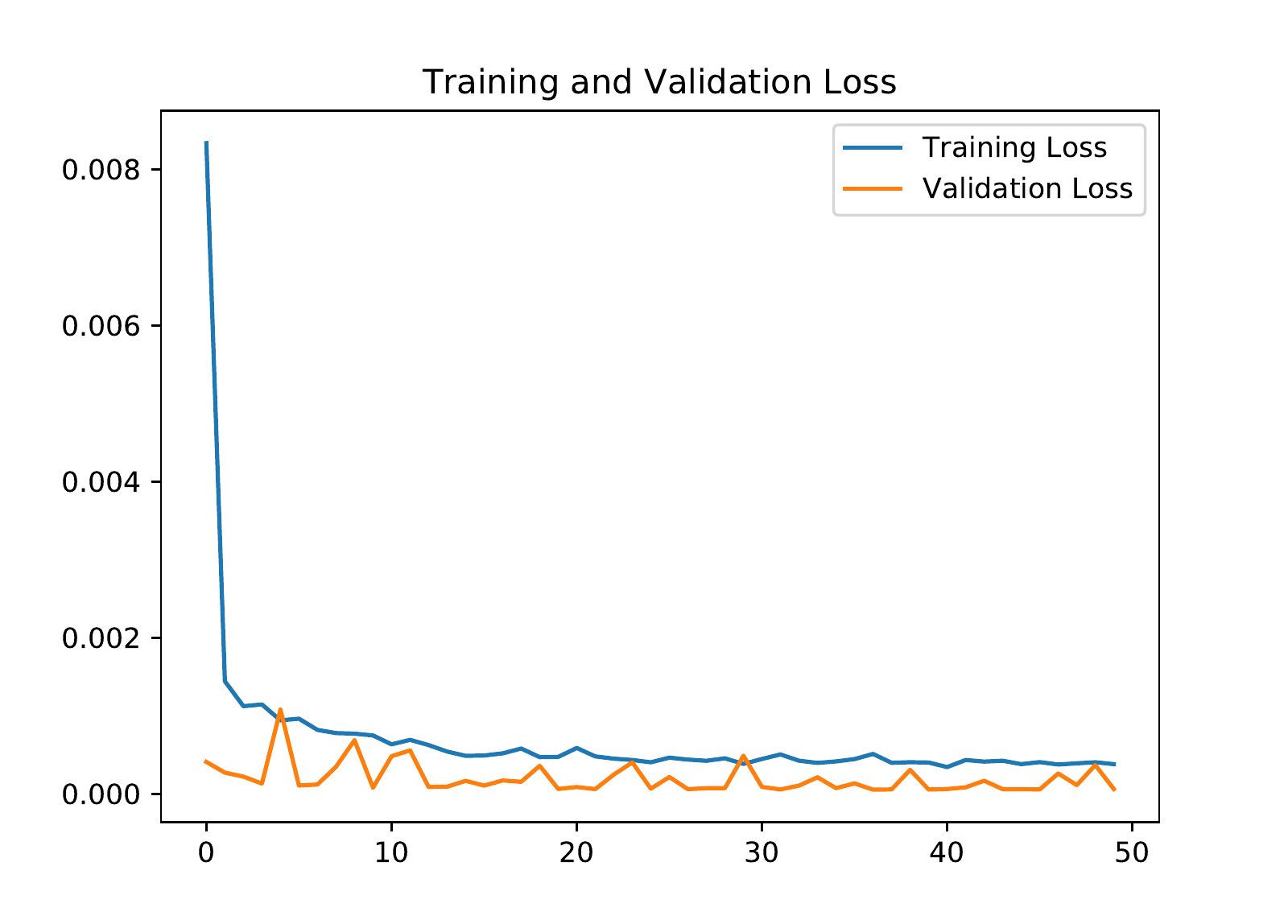}
    \caption{The loss curve of our proposed model.}\label{fig:lossa}
\end{figure}

The loss curve of our proposed model is shown on Fig. \ref{fig:lossa}.

\begin{figure}
    \centering
    \includegraphics[width=0.46\textwidth]{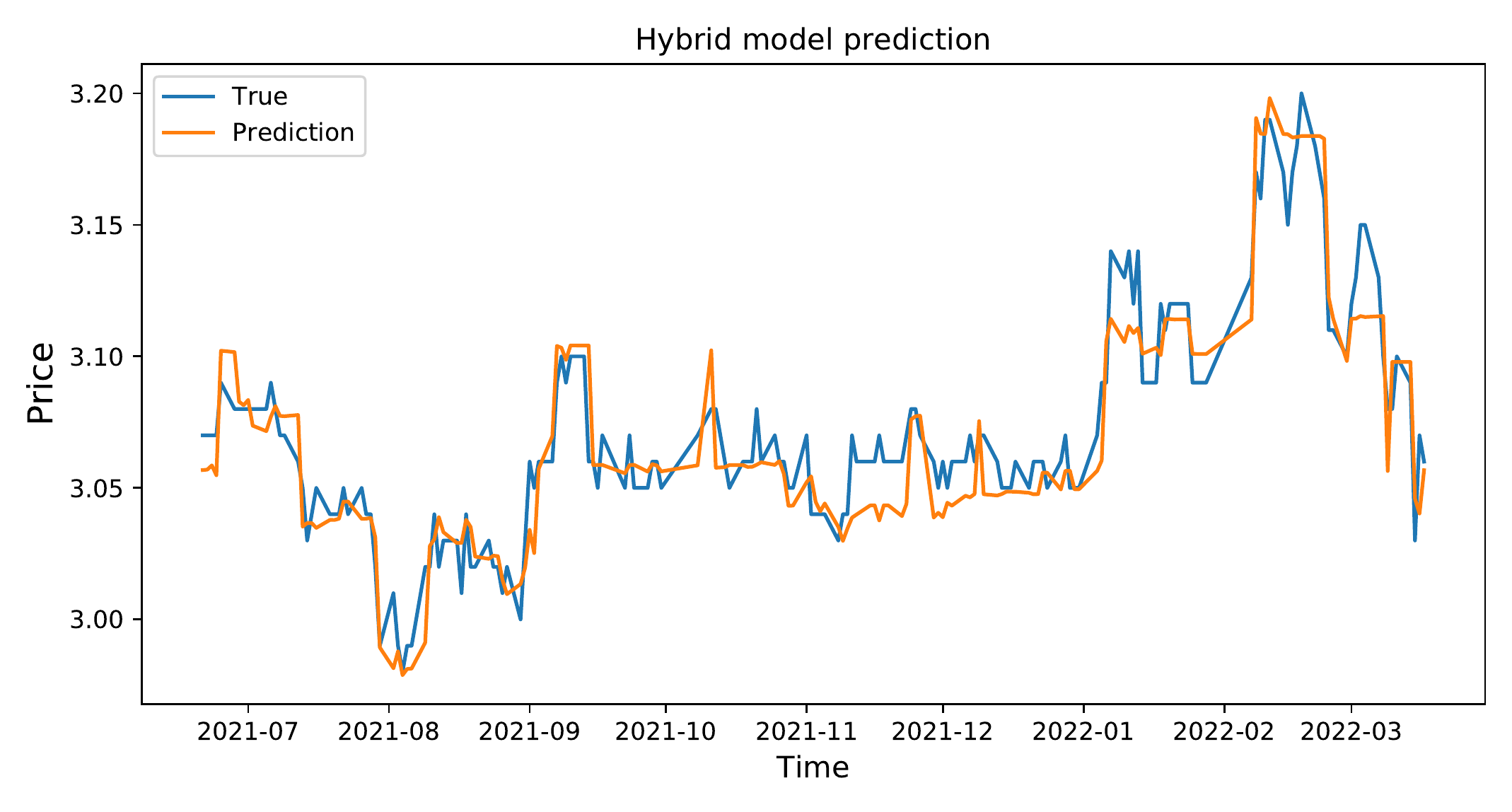}
    \caption{The stock price prediction result of our proposed model.}\label{fig:pred}
\end{figure}

The stock price prediction result of our proposed model is shown on Fig. \ref{fig:pred}.

\subsection{Compared with other methods}

The evaluation metrics are mean absolute error (MAE), root of mean square error (RMSE), mean absolute percentage error (MAPE) and $R^2$.
\begin{equation}
    MAE = \frac{1}{n}\sum\limits_{t = 1}^n {\left| {{{\hat X}_t} - {X_t}} \right|} 
\end{equation}
\begin{equation}
    RMSE = \sqrt {\frac{1}{n}\sum\limits_{t = 1}^n {{{({{\hat X}_t} - {X_t})}^2}} }
\end{equation}
\begin{equation}
    MAPE = \frac{1}{n}\sum\limits_{t = 1}^n {\left| {\frac{{{{\hat X}_t} - {X_t}}}{{{X_t}}}} \right|} 
\end{equation}
\begin{equation}
    R^2=\frac{\sum\limits_{t = 1}^n \|\hat X_t-\bar X_t\|^2}{\sum\limits_{t = 1}^n \|X_t-\bar X_t\|^2}
\end{equation}
Here $\bar X_t$ denotes the mean value of $X_t$. Lower error and higher $R^2$ denote better performance.

First, we conduct comparison with different pre-training and fine-tuning models. Table \ref{tab:acnn} demonstrates that our proposed model outperforms other baselines. 

\begin{table}[]
    \caption{Comparison with different pre-training and fine-tuning models}
    \label{tab:acnn}
    \begin{tabular}{@{}cccccc@{}}
    \toprule
    Pre-training & Fine-tuning & MSE     & RMSE    & MAE     & R2      \\ \midrule
    None         & None        & 0.00057 & 0.02734 & 0.02368 & 0.74402 \\
    None         & XGBoost     & 0.00031 & 0.01755 & 0.01223 & 0.82405 \\
    SL-LSTM      & SL-LSTM     & 0.00045 & 0.02282 & 0.01960 & 0.79434 \\
    ML-LSTM      & ML-LSTM     & 0.00031 & 0.01720 & 0.01265 & 0.82351 \\
    BiLSTM       & BiLSTM      & 0.00027 & 0.01652 & 0.01201 & 0.84210 \\
    BiLSTM       & XGBoost     & 0.00024 & 0.01605 & 0.01187 & 0.86301 \\
    CNN-BiLSTM   & XGBoost     & 0.00022 & 0.01529 & 0.01145 & 0.87720 \\
    ACNN-BiLSTM  & XGBoost     & 0.00020 & 0.01424 & 0.01126 & 0.88342 \\ \bottomrule
    \end{tabular}
\end{table}

Then, we conduct comparison with current methods. The compared methods include:
\begin{itemize}
    \item ARIMA
    \item ARIMA-NN \cite{ZHANG2003159} improved time-series forecasting using a hybrid ARIMA and neural network model.
    \item LSTM-KF, Transformer-KF, TL-KF \cite{shi2021tlkf} proposed Kalman Filter, along with LSTM and Transformer for stock prediction.
\end{itemize}
Table \ref{tab:stock1} demonstrate that the proposed neural networks outperforms current methods.

\begin{table}[]
    \centering
    \caption{Results on different methods}
    \label{tab:stock1}
    \begin{tabular}{@{}ccccc@{}}
    \toprule
    Model       & MAE    & RMSE   & MAPE   & R2\\ \midrule
    ARIMA          & 0.00057 & 0.02734 & 0.02368 & 0.74402 \\
    ARIMA-NN       & 0.00052 & 0.02608 & 0.02350 & 0.75037\\
    LSTM-KF        & 0.00047 & 0.02381 & 0.02192 & 0.76245 \\
    Transformer-KF & 0.00037 & 0.01924 & 0.01525 & 0.80230\\
    TL-KF          & 0.00033 & 0.01656 & 0.01372 & 0.81923\\
    AttCLX     & 0.00020 & 0.01424 & 0.01126 & 0.88342 \\\bottomrule
    \end{tabular}
\end{table}

\section{Conclusions}

Stock market is of great importance in the financial and economic development. Due to the complex volatility of the stock market, the prediction on the tendency of the stock price, can secure the return for the investors. The traditional time series model ARIMA can not describe the nonlinearity in the stock prediction. As neural networks are with strong nonlinear modeling ability, this paper proposes an attention-based CNN-LSTM and XGBoost hybrid model to predict the stock price. The model in this paper integrates the ARIMA model, the Convolutional Neural Networks with Attention mechanism, the Long Short-Term Memory network, and XGBoost regressor in a non-linear relationship, and improves the prediction accuracy. The model can capture the information of the stock market in multiple periods. The stock data is first preprocessed through ARIMA. Then, the deep learning architecture formed in pretraining-finetuning framework is adopted. The pre-training model is the Attention-based CNN-LSTM model based on sequence-to-sequence framework. The model first uses attention-based multi-scale convolution to extract the deep features of the original stock data, and then uses the Long Short-Term Memory networks to mine the time series features. Finally, the XGBoost model is adopted for fine-tuning. The results show that the hybrid model is more effective, which can help investors or institutions to achieve the purpose of expanding return and avoiding risk.



\bibliographystyle{IEEETrans}
\bibliography{reference}


\vfill

\end{document}